\documentclass{aastex}

\usepackage{emulateapj5}
\usepackage{apjfonts}
\usepackage{natbib}
\usepackage{graphicx}
\usepackage{amsmath}

\newcommand{\ltsim}{\ensuremath{\lesssim}}
\newcommand{\Msun}      {\mbox{$\rm\,M_{\mathord\odot}$}}

\newcommand{\cygxthree}	{\mbox{Cyg~X--3}}
\newcommand{\xray}	{\mbox{X-ray}}
\newcommand{\xrays}	{\mbox{X-rays}}
\newcommand{\chandra}{\textit{Chandra}}
\newcommand{\rxte}{\textit{RXTE}}
\newcommand{\asec}	{\mbox{$^{\prime \prime}$}}
\newcommand{\amin}	{\mbox{$^{\prime}$}}
\newcommand{\aprx}	{\mbox{$\sim$}}
\newcommand{\lumin}     {\mbox{$\rm\,ergs\,s^{-1}$}}

\newcommand{\degree}	{\hbox{$^\circ$}}
\newcommand{\eflux}     {\mbox{$\rm\,ergs~cm^{-2}~s^{-1}$}}
\newcommand{\chisq}     {\mbox{$\rm\,\chi^2$}}
\newcommand{\extra}	{20$^h$ 32$^m$ 27$^s$.1}
\newcommand{\extdec}	{$+$40\degree\ 57\amin\ 33\asec.5}
\newcommand{\lsi}	{LS~I~$+$61\degree303}

\shorttitle{Extended Emission in \cygxthree}
\shortauthors{Heindl, et al.}

\begin{document}
\title{Extended Emission from Cygnus~X$-$3 Detected with \chandra} 
\author{W.A. Heindl, J.A. Tomsick}
\affil{Center for Astrophysics and Space Sciences, Code 0424, University of
California, San Diego, La Jolla, CA 92093}
\author{R. Wijnands}
\affil{School of Physics and Astronomy, University of St Andrews,
  St Andrews, Fife, KY16 9SS, Scotland, UK
}
\author{D.M. Smith}
\affil{Space Sciences Laboratory, University of California,
Berkeley, Centennial at Grizzly Peak Boulevard, Berkeley,
CA 94720-7450}

\email{wheindl@ucsd.edu}

\begin{abstract}


We have discovered extended \xray\ emission from the microquasar
\cygxthree\ in archival \emph{Chandra X-ray Observatory} observations.
A 5\asec\ wide structure lies approximately 16\asec\ to the NE from
the core point source and may be extended in that direction.  This
angular scale corresponds to a physical extent of roughly 0.8\,lyr, at
a distance of 2.5\,lyr from \cygxthree\ (assuming a 10\,kpc
distance). The flux varied by a factor of 2.5 during the four months
separating two of the observations, indicating significant
substructure.  The peak 2--10\,keV luminosity was
\aprx$5\times10^{34}$\,\lumin. There may also be weaker, extended
emission of similar scale oppositely directed from the core,
suggesting a bipolar outflow.  This structure is not part of the dust
scattering halo, nor is it caused by the \chandra\ point spread
function.  In this \textit{Letter} we describe the observations and
discuss possible origins of the extension.

\end{abstract}


\section{Introduction}

\cygxthree\ is a bright high mass \xray\ binary with a 4.8\,hr orbital
period lying at a distance of \aprx10\,kpc \citep{Dick83}. The exact
nature (black hole or neutron star) of the compact object is still in
question. The mass donating companion is most likely an early
Wolf-Rayet star \citep{Fen99} with an intense stellar wind evidenced
by both infrared \citep{vanKerk93} and \xray\ \citep{Pae00} lines.
The source is highly variable from radio to hard \xray\ wavelengths.
In the radio, several Jansky flares are associated with the ejection
of relativistic radio jets \citep[see e.g., ][]{Gel83,Mio01}.
Depending on assumptions made for the jet geometry, intrinsic
velocities between 0.3c -- 0.8c have been inferred. \cygxthree\ is
therefore one of about a dozen Galactic binaries considered to be
``microquasars'' \citep[see ][ for a list]{Dis02}.  When observed at
milliarcsecond scales, the radio jets appear to be one-sided, while on
arcsecond scales they appear two-sided.  In either case the jet axis
is nearly north-south.

\section{Observations and Analysis \label{s:obs}}

\begin{figure}
\centerline{
\includegraphics[height=6.25in,angle=90,bb=64 85 407 713,clip]{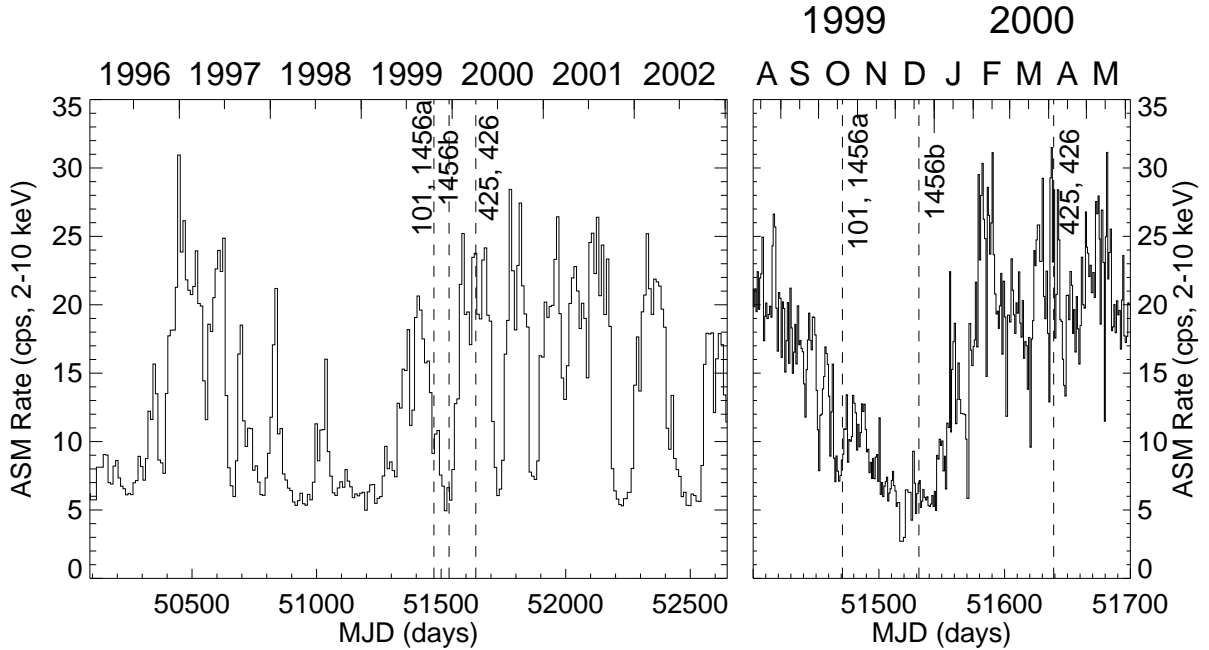}}
\caption{\label{f:asm} The \rxte /ASM light curve of Cyg~X-3. Left:
full light curve, averaged in 10\,day bins.  Right: one day averages
in region surrounding the \chandra\ observations.  The dates of the \chandra/HETG
observations are indicated.  }
\end{figure}

\chandra\ has observed \cygxthree\ with the spectroscopy array of the 
Advanced CCD Imaging Spectrometer/High Energy Transmission Grating
(ACIS-S/HETG) on four occasions for the purposes of studying the
Wolf-Rayet wind \citep{Pae00}, measuring the source distance
\citep{Pre00}, and searching for emission lines from the radio
jets. The observations were made in 1999 October and December and in
2000 April (see Table~\ref{t:obs}). Figure~\ref{f:asm} puts the
\chandra\ observations in the context of the \emph{Rossi X-ray Timing
Explorer} All Sky Monitor \mbox{(\rxte /ASM)} Cyg~X-3 light curve. The
\xray\ flux exhibits a strong bimodal behavior,
alternating between extended periods below 10 and above
20\,cps. Observations 101+1456a and 1456b occurred during a transition
to and in a low state, respectively, while 425 and 426 were during the
high state.  Since the scattering halo is proportional to the recent
point source flux, the high state observations have much brighter halo
emission which is effectively a background for the detection of
extended structure.

In our analysis of these observations, we used the CIAO tools version
2.2.  The archival data were processed with ASCDS versions R4CU5UPD9
(101, 1456) and 6.3.1 (425, 426).

\subsection{Image Analysis}

\begin{figure}
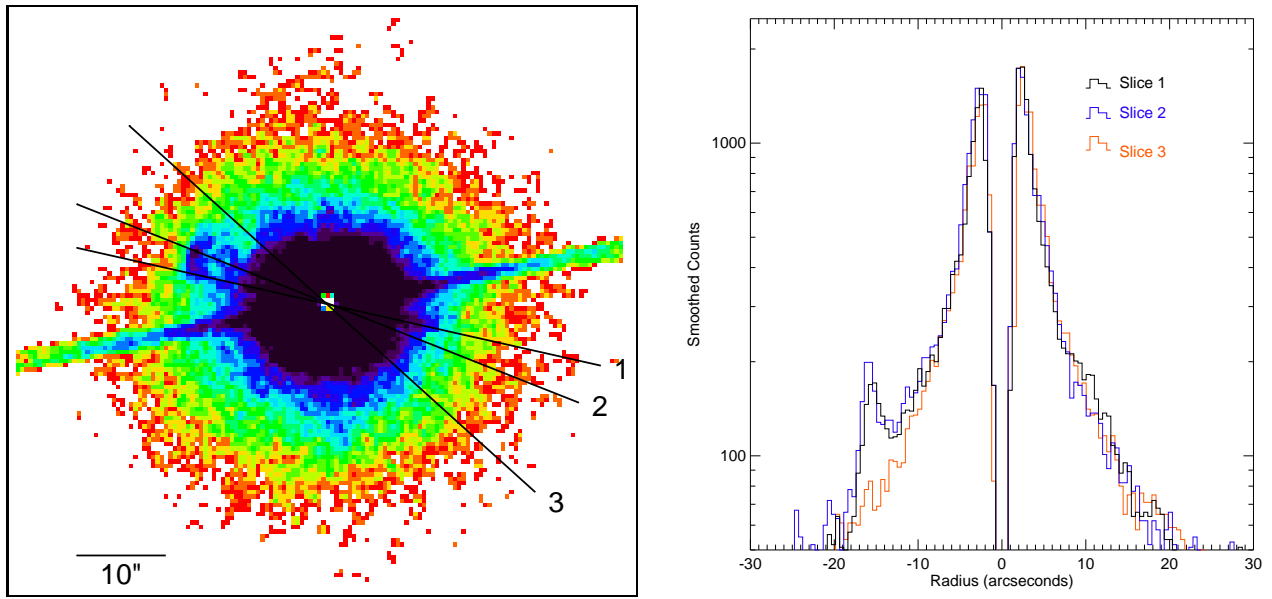

\centerline{\fbox{\includegraphics[angle=0,height=3in,bb=66 61 453 423,clip]{cygx3_image.cps}}\hspace*{0.25in}\includegraphics[height=3in,bb=85 83 528 523]{comp_slices.cps} }
\caption{\label{f:image} Left: Total zero order image of all four
observations smoothed with a 2$\times$2 boxcar function. Image is
69\asec\ square. North is up, and East is to the left. The core of the
PSF is heavily piled up, and the bulk of the surrounding emission is
the dust scattering halo \citep{Pre00}. The narrow SE to NW feature is
the ACIS readout strip from the 2000 observations when Cyg~X-3 was
bright.  Excess, extended emission is visible to the NE and perhaps
the SW of the core. Right: Profiles of the image along the three
indicated slices. In slices 1 and 2, the excess emission is clearly
visible about 16\asec\ NW from the core.  In slice 1, there may also
be excess emission \aprx8\asec--13\asec\ to the SW of the core,
suggesting oppositely directed outflows. Slice 3 avoids the extensions
and is shown for comparison.}
\end{figure}

For each observation, we extracted the zeroth order image binned at
the nominal ACIS-S resolution (\aprx0.49\asec).  Figure~\ref{f:image}
shows the total image from all four observations. In each individual
image, there is a bright, unresolved core surrounded by a strong
scattering halo \citep{Pre00} and the point spread function (PSF)
wings. In each case, there is also significant (\aprx10$\sigma$)
excess emission to the NE (position angle, PA\aprx75\degree) of the
point source. The farthest extent from the core is 18\asec,
corresponding to 2.7\,$D_{10}$\,lyr where $D_{10}$ is the distance in
units of 10\,kpc. It is also clearly extended, by about 5\asec\
(0.8\,$D_{10}$\,lyr), orthogonal to the core direction.  As a fiducial
location for the extension, its brightest point, near the center of
the outermost arc, lies at RA=\extra, Dec=\extdec\ (J2000).  We are
confident the extension is real, as it appears at the same sky
location in each observation, even though the observations were made
at very different roll angles (see Table~\ref{t:obs}). This rules out
instrumental artifacts, such as PSF asymmetries, which would move with
the roll angle. Figure~\ref{f:image} also shows the profile of three
slices through the point source.  Although the point source is piled
up, resulting in lost counts on axis, a clear extension is seen
\aprx16\asec\ off axis in two of the slices. The third, which avoids
the extension, is shown for comparison. Furthermore, there is some
evidence for a SW extension oppositely directed along slice 1.
Neither feature is aligned with the ACIS readout streak or dispersed
grating spectra in any of the observations.

It is not clear whether the NE extension is connected to the point
source or not. Apparent in the image is a ``bridge'' of emission
stretching from the NE extension toward the point source. The two
slices suggest that this bridge could simply be the scattering halo,
which shows modest azimuthal variations at most radii. However, if
this bridge is associated with the extension, then there is evidence
of substructure, since it appears brighter around the periphery with
relatively weak emission in the center.

Finally, the milliarcsecond scale radio jets are almost
north-south aligned, with PA=175\degree\ \citep{Mio01}, nearly, but not
quite, perpendicular to the \xray\ extension.

\subsection{Spectra} 

In order to search for any connection between the point source
emission and that of the NE extension, we extracted grating spectra of
the point source and CCD spectra of the extension from all four
observations.  This allowed us to look for correlated time variability
and spectral similarities. For the extended emission, we used segments
of annuli taken at the same radial distance from the point source for
source and background regions.  Table~\ref{t:rates} lists rates,
fluxes, and spectral parameters from fits to absorbed black body and
power law models. For the point source, which has strong, variable
lines from the stellar wind \citep{Pae00}, we included the six or
seven most prominent lines as required for each observation in order
to achieve an acceptable chi-squared.  Good fits were achieved with
both models. The resulting \chisq\ values gave no preference between
the two, except for the core source in observations 425 and 426 where
a black body was somewhat preferred.

Within each observation, the extension flux was consistent with a
constant. The flux was also unchanged between observations 425 and
426, which were separated by two days. We therefore have no evidence
of variability on less than \aprx day times.  However, the point
source and scattering halo did brighten by a factor of seven between
observations 1456b and 425 (separated by four months), while the
extension count rate only increased by about a factor of 2.5. This
implies that the emission is not simply reprocessed/scattered point
source flux (see \S\ref{s:disc}).  In any event, given the apparent
size (\aprx0.8\,lyr) of the extended region, it is somewhat surprising
to observe such strong variability in just four months. This hints
that there may be significant underlying substructure. Unfortunately,
because of limited statistics, we cannot determine if the extension
spectrum changed, although the 1999 spectra are marginally harder than
those in 2000.

\section{Discussion \label{s:disc}}

Any model of the extended emission must explain the observed
morphology, time variability, luminosity, and spectrum.  We consider
several possible origins: scattering in a non-axisymmetric dust halo,
a coincidental alignment of an unrelated source, reflection from the
companion's stellar wind, bremsstrahlung from the disk wind, and
collision of historical jets with the ISM.

One possibility is that a non-uniform distribution of dust around the
line of sight could result in enhanced, asymmetric halo emission.  In
this case, the spectrum and flux should follow closely the point
source spectrum with only a short (\ltsim days) delay.  However, the
extension flux varied only by a factor of 2.5 over the observations,
while the point source and the bulk of the halo varied by a factor of
seven.  Also, the black body temperature of the point source and
extension in ObsID 425 (when \cygxthree\ had been bright for weeks)
are significantly different (see Table~\ref{t:rates}). These facts
argue that scattering halo asymmetries are not the cause of the
extension.

We also considered the possibility that the source is unrelated to
\cygxthree.  However, the fact that it is both extended and variable
on month timescales rules out any extra-galactic object. The only
remaining possibilities are supernova remnants, pulsar wind nebulae, and
microquasar jets.  Supernova remnants are ruled out by the time
variability. The lack of a radio counterpart and the low number
density of pulsar nebulae \citep[only about 25 are
known;][]{Helfand98} make this possibility unlikely,
P\aprx$10^{-8}$. Finally any microquasar jets would certainly be
associated with \cygxthree.  We therefore think it very unlikely that
this is a coincidental alignment.

The emission may be flux from the point source scattered by the
integrated Wolf-Rayet wind well outside the binary system.  However,
if the wind is persistent, then the outer regions of the disk wind
should be shielded by the denser inner regions.  This would result in
a morphology dominated by flux close to the system, very different
from what we observe.

However, the Wolf-Rayet wind may still cause the extension via
Bremsstrahlung at its interface with the ISM. \citet{Ogl01} used {\em
Infrared Space Observatory} observations to infer a mass loss rate,
which for disk-like wind geometries subtending 1.2\,sr, could be as
high as $10^{-3}$\Msun\,$\rm yr^{-1}sr^{-1}$.  Such a disk wind in the
plane of the binary orbit was suggested by \citet{Fen99} to explain
variations in the infrared spectrum and orbital modulation of
V(iolet)/R(ed). Furthermore, \citet{Mar01} hypothesize that a disk
wind inclined to the radio jet axis could obscure one side of the base
of a double jet. This could cause the jet to appear one-sided on
milliarcsecond scales and two-sided on larger scales, as is observed.
With a terminal velocity of \aprx1500\,km\,s$^{-1}$ \citep{vanKerk93},
such a wind would require about 500\,$D_{10}$\,years to reach the
observed 18\asec\ extent of the \xray\ emission. This would
effectively be a minimum age for the observed extension.  A
Bremsstrahlung fit to the ObsID 425 spectrum gives an electron
temperature of $1.6^{+1.4}_{-0.6}$\,keV. The other observations were
poorly constrained, but consistent with ObsID 425. With this
temperature, and assuming a volume corresponding to a 5\asec\ diameter
sphere, the data imply an electron density of a few hundred per cubic
centimeter. While this is much higher than even the densest disk winds
at a radius of 2.5\,lyr, if the wind is plowing into the ISM and
stopping, then the pertinent quantity is the \emph{time integrated}
wind density. Taking the above disk wind parameters and assuming the
wind is fully ionized, it would require only about 2,000 years to
build up the necessary electron density. Time variability would result
from (historical) changes in the wind intensity that would heat the
accumulated material to a greater or lesser degree.  This picture is
our most consistent explanation for the observations.

The emission might also be the result of material from an earlier jet
ejection colliding with the ISM.  In this case the \xrays\ would most
likely be from synchrotron emission, as appears to be the case for the
\xray\ jets in XTE~J1550-564 \citep{Tomsick03}. This picture would account
for the possible limb brightening of the extension, as the bulk of the
emission would be expected at the jet/ISM interface.  However, this
scenario would require that the jet direction has precessed through a
very large angle, since the observed radio jets are aligned close to
the north-south direction. Furthermore, the transverse extent of the
\xray\ structure requires that the jet was either poorly 
collimated or exhibited significant change in direction during the
emission period.  This is significantly different from XTE~J1550-564
where the jets are collimated to 2\degree\citep{Corbel02}. We note that the
morphology expected from jet emission, a conical feature emanating
from the core, should be easily distinguished from the elliptical
shape expected from the disk wind model. We have recently made a 
\chandra\ High Resolution Camera observation to try to distinguish
these possibilities.

Finally, we note that the microquasars SS433, \lsi, and possibly
Cyg~X-1 have extended \emph{radio} emission around their cores
perpendicular to their jet axes
\citep{Paragi99,Massi01,Stirling98,Spencer01}.  While these features
are on much smaller (AU) scales, it is suggested that they are related
to disk-like winds in the binary orbital plane.  This suggests that
the extended \xray\ regions may be visible as radio sources, and deep
searches should be carried out.

\bibliographystyle{astron_noskip}



\normalsize

\begin{table}
\begin{minipage}{\textwidth}
\renewcommand{\thefootnote}{\thempfootnote}
\caption{\label{t:obs}\chandra\ ACIS-S/HETG observations of \cygxthree.}
\begin{tabular}{llccc}
\hline\hline
 		& 		& Exposure	& Roll Angle	& ASM
Rate\footnote{\rxte/ASM average rate (2--10\,keV) on the date of observation.}\\ 
ObsID\footnote{Observation Identification}		& Date	\footnote{UT at mid-point of observation.}	& (ks)		& (deg)		& (cps)	\\\hline
101$+$1456a\footnote{ObsID 1456 was split in two ``observation
intervals'' with different roll angles which
we designate 1456a and 1456b separated by two months. ObsID 101
was contiguous with 1456a and we analyzed their spectra jointly.}	& 1999 Oct 20.1	& 2$+$12	& 269		& 9.1 \\
1456b		& 1999 Dec 19.3 & 8.4		& 322		& 6.9 \\
425		& 2000 Apr 4.7	& 16		& 78		& 23 \\
426		& 2000 Apr 6.9	& 14		& 81		& 18 \\\hline
	
\end{tabular}
\end{minipage}
\end{table}

\begin{table}
\begin{minipage}{\textwidth}
\renewcommand{\thefootnote}{\thempfootnote}
\caption{\label{t:rates}Spectral fits to NE extension and core point
source. Uncertainties are 90\% confidence for one parameter of
interest.}  
\footnotesize
\begin{tabular}{rccccc} \hline \hline
ObsID & \multicolumn{2}{c}{NE Extension} && \multicolumn{2}{c}{Core} \\ 
\cline{2-3} \cline{5-6}
 & Black Body & Power Law && Black Body & Power Law \\ \hline 
\multicolumn{1}{l}{101$+$1456a}  & & && & \\
\cline{1-1}
 Rate\footnote{NE extension: HETG order 0. Point source: HEG+1. counts/s} & 
	\multicolumn{2}{c}{$ (2.19\pm0.20)\times 10^{-2} $} & &  \multicolumn{2}{c}{$ 2.94\pm0.02 $}  \\
 N$_H$ & $2.6^{+4.0}_{-2.6}$ & $4.6^{+7.1}_{-4.1}$ && $2.94^{+0.02}_{-0.02}$ & $4.10^{+0.16}_{-0.16}$\\
 kT (keV)/$\Gamma$ & $2.2^{+1.9}_{-0.8}$ & $0.8^{+1.5}_{-1.1}$ && $1.88^{+0.04}_{-0.04}$ & $1.07^{+0.05}_{-0.05}$ \\
 Flux \footnote{unabsorbed, \eflux, 2--10\,keV}& $2.0\times 10^{-12} $& $2.8\times 10^{-12}$&& $2.1\times 10^{-9}$& $2.6\times 10^{-9}$\\
 $\chi^2$/dof & 0.71/64 & 0.70/64&& 0.84/2042 & 0.85/2042 \\ \hline 
\multicolumn{1}{l}{1456b} & & && & \\
\cline{1-1}
 Rate & \multicolumn{2}{c}{$(1.56 \pm 0.20)\times 10^{-2}$} & &  \multicolumn{2}{c}{$ 1.91\pm0.02 $}  \\
 N$_H$ & $1.2^{+6.5}_{-1.2}$ & $3.9^{+9.7}_{-3.9}$ && $2.15^{+0.21}_{-0.21}$ &  $3.9^{+0.3}_{-0.3}$\\
 kT (keV)/$\Gamma$ & $2.0^{+3.5}_{-0.9}$ & $1.0^{+2.2}_{-1.4}$ && $2.20^{+0.10}_{-0.10}$ & $0.74^{+0.09}_{-0.09}$ \\
 Flux & $1.4\times 10^{-12}$ &  $1.7\times 10^{-12}$ && $1.4\times 10^{-9}$ & $1.9\times 10^{-9}$ \\
 $\chi^2$/dof & 0.96/25 & 0.95/25 &&0.89/1013& 0.91/1013  \\ \hline 
\multicolumn{1}{l}{425}  & & && & \\
\cline{1-1}
 Rate & \multicolumn{2}{c}{$ (3.86\pm 0.34)\times 10^{-2}$} & &  \multicolumn{2}{c}{$13.38\pm 0.03 $}  \\
 N$_H$ & $6.2^{+5.8}_{-3.8}$ & $13.2^{+8.2}_{-6.1}$ && $3.93^{+0.04}_{-0.04}$ & $7.11^{+0.08}_{-0.05}$ \\
 kT (keV)/$\Gamma$ & $0.80^{+0.21}_{-0.21}$ & $4.7^{+2.0}_{-1.6}$ && $1.31^{+0.01}_{-0.01}$ & $2.17^{+0.02}_{-0.02}$ \\
 Flux &$5.3\times 10^{-12}$ & $1.9\times 10^{-11}$ && $1.1\times 10^{-8}$  &  $1.3\times 10^{-8}$\\
 $\chi^2$/dof & 1.24/23 &1.3/23 && 1.33/3828 & 1.67/3828 \\ \hline 
\multicolumn{1}{l}{426}  & & && & \\
\cline{1-1}
 Rate & \multicolumn{2}{c}{$ (3.7\pm0.3) \times 10^{-2}$} & &  \multicolumn{2}{c}{$ 10.34 \pm 0.03 $}  \\
 N$_H$ & $6.9^{+7.1}_{-5.7}$ & $14.1^{+10.2}_{-8.3}$ && $5.57^{+0.08}_{-0.08}$ &  $9.8^{+0.1}_{-0.1}$\\
 kT (keV)/$\Gamma$ & $1.0^{+0.4}_{-0.3}$ & $3.8^{+2.2}_{-1.8}$ && $1.32^{+0.01}_{-0.01}$ & $2.36^{+0.03}_{-0.03}$ \\
 Flux &$4.7\times 10^{-12}$ & $1.3\times 10^{-11}$ && $8.5\times 10^{-9}$ & $1.4\times 10^{-8}$ \\
 $\chi^2$/dof & 1.27/16 & 1.28/16 && 1.12/3068 & 1.34/3068 \\ 
\hline \hline
\normalsize
\end{tabular}
\end{minipage}
\end{table}

\end{document}